\documentclass[prd,aps,showpacs,epsf,floats,10pt]{revtex4}
\usepackage{amssymb}
\usepackage{amsfonts}
\usepackage{amsmath}

\input{tcilatex}
\begin{document}

\title{The Effect Of Microscopic Correlations On The Information Geometric
Complexity Of Gaussian Statistical Models}
\author{S. A. Ali$^{1}$, C. Cafaro$^{2}$, D.-H. Kim$^{3}$, S. Mancini$^{4}$}
\affiliation{$^{1}$Department of Physics, State University of New York at Albany, 1400
Washington Avenue, Albany, NY 12222, USA\\
$^{2\text{, }4}$Dipartimento di Fisica, Universit\`{a} di Camerino, I-62032
Camerino, Italy\\
$^{3}$Institute for the Early Universe, Ewha Womans University, Daehyun-dong
11-1, Seodaemun-gu, Seoul 120-750, South Korea and Center for Quantum
Spacetime, Sogang University, Shinsu-dong 1, Mapo-gu, Seoul 121-742, South
Korea }

\begin{abstract}
We present an analytical computation of the asymptotic temporal behavior of
the information geometric complexity (IGC) of finite-dimensional Gaussian
statistical manifolds in the presence of microcorrelations (correlations
between microvariables). We observe a power law decay of the IGC at a rate
determined by the correlation coefficient. It is found that
microcorrelations lead to the emergence of an asymptotic information
geometric compression of the statistical macrostates explored by the system
at a faster rate than that observed in absence of microcorrelations. This
finding uncovers an important connection between (micro)-correlations and
(macro)-complexity in Gaussian statistical dynamical systems.
\end{abstract}

\maketitle

\section{\textbf{Introduction}}

The study of complexity \cite{gell-mann} has created a new set of ideas on
how very simple systems may give rise to very complex behaviors. In many
cases, the "laws of complexity" have been found to hold universally,
independent of the details of the system's constituents. Chaotic behavior is
a particular case of complex behavior and it will be the object of the
present work. In this article we make use of the so-called Entropic Dynamics
(ED) \cite{caticha1} and Information Geometrodynamical Approach to Chaos
(IGAC) \cite{carlo-tesi, carlo-CSF}. ED arises from the combination of
inductive inference (Maximum Entropy Methods, \cite{caticha2, adom1}) and
Information Geometry \cite{amari}. ED is a theoretical framework whose
objective - among others - is to derive dynamics from purely entropic
arguments. The applicability of ED has been extended to temporally-complex
(chaotic) dynamical systems on curved statistical manifolds $\mathcal{M}_{S}$
resulting in the information geometrodynamical approach to chaos (IGAC) \cite%
{carlo-tesi}. IGAC arises as a theoretical framework to study chaos in
informational geodesic flows describing physical, biological or chemical \
systems. A geodesic on a curved statistical manifold $\mathcal{M}_{S}$
represents the maximum probability path a complex dynamical system explores
in its evolution between initial and final macrostates. Each point of the
geodesic is parametrized by the macroscopic dynamical variables\textbf{\ }$%
\left\{ \Theta \right\} $ defining the macrostate of the system.
Furthermore, each macrostate is in a one-to-one correspondence with the
probability distribution $\left\{ p\left( X|\Theta \right) \right\} $
representing the maximally probable description of the system being
considered. The set of macrostates forms the parameter space $\mathcal{D}%
_{\Theta }$ while the set of probability distributions forms the statistical
manifold $\mathcal{M}_{S}$. IGAC\ is the information geometric analogue of
conventional geometrodynamical approaches \cite{casetti, di bari} where the
classical configuration space $\Gamma _{E}$\ is being replaced by a
statistical manifold $\mathcal{M}_{S}$. This procedure affords the
possibility of considering chaotic dynamics arising from non conformally
flat metrics (the Jacobi metric is always conformally flat, instead). It is
an information geometric extension of the Jacobi geometrodynamics (the
geometrization of a Hamiltonian system by transforming it to a geodesic flow 
\cite{jacobi}). The reformulation of dynamics in terms of a geodesic problem
allows the application of a wide range of well-known geometrical techniques
in the investigation of the solution space and properties of the equation of
motion. The power of the Jacobi reformulation is that all of the dynamical
information is collected into a single geometric object in which all the
available manifest symmetries are retained- the manifold on which geodesic
flow is induced. For example, integrability of the system is connected with
existence of Killing vectors and tensors on this manifold. The sensitive
dependence of trajectories on initial conditions, which is a key ingredient
of chaos, can be investigated from the equation of geodesic deviation. In
the Riemannian \cite{casetti} and Finslerian \cite{di bari} (a Finsler
metric is obtained from a Riemannian metric by relaxing the requirement that
the metric be quadratic on each tangent space) geometrodynamical approach to
chaos in classical Hamiltonian systems, an active field of research concerns
the possibility of finding a rigorous relation among the sectional
curvature, the Lyapunov exponents, and the Kolmogorov-Sinai dynamical
entropy (i. e. the sum of positive Lyapunov exponents) \cite{kawabe}.

Understanding the relationship between microscopic dynamics and
experimentally observable macroscopic dynamics is a fundamental issue in
physics \cite{leb81, leb93, leb99, tom90}. An interesting manifestation of
such a relationship appears in the study of the effects of microscopic
external noise (noise imposed on the microscopic variables of the system) on
the observed collective motion (macroscopic variables) of a globally coupled
map \cite{shib99}. These effects are quantified in terms of the complexity
of the collective motion. Furthermore, it turns out that noise at a
microscopic level reduces the complexity of the macroscopic motion, which in
turn is characterized by the number of effective degrees of freedom of the
system.

In this article, using statistical inference and information geometric
techniques, we investigate the macroscopic behavior of complex systems in
terms of the underlying statistical structure of its microscopic degrees of
freedom in the presence of correlations. We compute the asymptotic temporal
behavior of the information geometric complexity of the maximum probability
trajectories on finite-dimensional Gaussian statistical manifolds in the
presence of microcorrelations. We observe a power law decay of the IGC at a
rate determined by the correlation coefficient. The ratio between the IGC in
the presence and in the absence of microcorrelations is explicitly computed.
We conclude that microcorrelations lead to the emergence of an asymptotic
information geometric compression of the explored statistical macrostates
(on the configuration manifold of the model in its evolution between the
initial and final macrostates) that is faster than that observed in absence
of microcorrelations.

The layout of the article is as follows. In Section II, we briefly discuss
Gaussian statistical models in absence and presence of microcorrelations. In
Section III, we introduce the Gaussian statistical model being considered.
We compute the Ricci scalar curvature and the geodesic trajectories of the
system. In Section IV, we compute the asymptotic temporal behavior of the
dynamical IGC of the model. Our conclusions are presented in Section V.

\section{On Gaussian Statistical Models and Microcorrelations}

In this Section, we introduce the notion of Gaussian statistical models
(manifolds) in the presence of correlations between the microscopic degrees
of freedom (microvariables) of the system (microcorrelations).

\subsection{Statistical Models in Absence of Microcorrelations}

Consider a Gaussian statistical model whose microstates span a $n$%
-dimensional space labelled by the variables $\left\{ X\right\} =\left\{
x_{1}\text{, }x_{2}\text{,..., }x_{n}\right\} $ with $x_{j}\in 
\mathbb{R}
$, $\forall j=1$,..., $n$. We assume the only testable information
pertaining to the quantities $x_{j}$ consists of the expectation values $%
\left\langle x_{j}\right\rangle $ and the variance $\Delta x_{j}$. The set
of these expected values define the $2n$-dimensional space of macrostates of
the system. A measure of distinguishability among the macrostates of the
Gaussian model is achieved by assigning a probability distribution $P\left(
X|\Theta \right) $ to each $2n$-dimensional macrostate $\Theta \overset{%
\text{def}}{=}\left\{ \left( ^{\left( 1\right) }\theta _{j}\text{,}^{\left(
2\right) }\theta _{j}\right) \right\} _{n\text{-pairs}}$ $=\left\{ \left(
\left\langle x_{j}\right\rangle \text{, }\Delta x_{j}\right) \right\} _{n%
\text{-pairs}}$. The process of assigning a probability distribution to each
state endows $\mathcal{M}_{S}$ with a metric structure. Specifically, the
Fisher-Rao information metric $g_{\mu \nu }\left( \Theta \right) $ \cite%
{amari} is a measure of distinguishability among macrostates on the
statistical manifold $\mathcal{M}_{S}$, 
\begin{equation}
g_{\mu \nu }\left( \Theta \right) =\int dXP\left( X|\Theta \right) \partial
_{\mu }\log P\left( X|\Theta \right) \partial _{\nu }\log P\left( X|\Theta
\right) =4\int dX\partial _{\mu }\sqrt{P\left( X|\Theta \right) }\partial
_{\nu }\sqrt{P\left( X|\Theta \right) }\text{,}  \label{FRM}
\end{equation}%
with $\mu $, $\nu =1$,..., $2n$ and $\partial _{\mu }=\frac{\partial }{%
\partial \Theta ^{\mu }}$. It assigns an information geometry to the space
of states. The information metric $g_{\mu \nu }\left( \Theta \right) $ is a
symmetric, positive definite Riemannian metric. For the sake of completeness
and in view of its potential relevance in the study of correlations, we
point out that the Fisher-Rao metric satisfies the following two properties:
1) invariance under (invertible) transformations of microvariables $\left\{
x\right\} \in \mathcal{X}$\textbf{;} 2) covariance under reparametrization
of the statistical macrospace $\left\{ \theta \right\} \in \mathcal{D}%
_{\theta }$. The invariance of $g_{\mu \nu }\left( \theta \right) $ under
reparametrization of the microspace $\mathcal{X}$ implies \cite{amari}%
\textbf{,}%
\begin{equation}
\mathcal{X}\subseteq 
\mathbb{R}
^{n}\ni x\longmapsto y\overset{\text{def}}{=}f\left( x\right) \in \mathcal{Y}%
\subseteq 
\mathbb{R}
^{n}\Longrightarrow p\left( x|\theta \right) \longmapsto p^{\prime }\left(
y|\theta \right) =\left[ \frac{1}{\left\vert \frac{\partial f}{\partial x}%
\right\vert }p\left( x|\theta \right) \right] _{x=f^{-1}\left( y\right) }%
\text{.}
\end{equation}%
The covariance under reparametrization of the parameter space\textbf{\ }$%
\mathcal{D}_{\theta }$\textbf{\ }(homeomorphic to\textbf{\ }$\mathcal{M}_{S}$%
\textbf{) }implies \cite{amari},%
\begin{equation}
\mathcal{D}_{\theta }\ni \theta \longmapsto \theta ^{\prime }\overset{\text{%
def}}{=}f\left( \theta \right) \in \mathcal{D}_{\theta ^{\prime
}}\Longrightarrow g_{\mu \nu }\left( \theta \right) \longmapsto g_{\mu \nu
}^{\prime }\left( \theta ^{\prime }\right) =\left[ \frac{\partial \theta
^{\alpha }}{\partial \theta ^{\prime \mu }}\frac{\partial \theta ^{\beta }}{%
\partial \theta ^{\prime \nu }}g_{\alpha \beta }\left( \theta \right) \right]
_{\theta =f^{-1}\left( \theta ^{\prime }\right) }\text{,}
\end{equation}%
where%
\begin{equation}
g_{\mu \nu }^{\prime }\left( \theta ^{\prime }\right) =\int dxp^{\prime
}\left( x|\theta ^{\prime }\right) \partial _{\mu }^{\prime }\log p^{\prime
}\left( x|\theta ^{\prime }\right) \partial _{\nu }\log p^{\prime }\left(
x|\theta ^{\prime }\right) \text{,}
\end{equation}%
with $\partial _{\mu }^{\prime }=\frac{\partial }{\partial \theta ^{\prime
\mu }}$ and $p^{\prime }\left( x|\theta ^{\prime }\right) =p\left( x|\theta
=f^{-1}\left( \theta ^{\prime }\right) \right) $. Our $2n$\textbf{-}%
dimensional Gaussian statistical model represents a macroscopic
(probabilistic) description of a microscopic $n$-dimensional (microscopic)
physical system evolving over\textbf{\ }a $n$-dimensional (micro) space.\
The variables $\left\{ X\right\} =\left\{ x_{1}\text{, }x_{2}\text{,..., }%
x_{n}\right\} $ label the $n$-dimensional space of microstates of the
system. We assume that all information relevant to the dynamical evolution
of the system is contained in the probability distributions. For this
reason, no other information is required. Each macrostate may be thought of
as a point of a $2n$-dimensional statistical manifold with coordinates given
by the numerical values of the expectations $^{\left( 1\right) }\theta _{j}$
and $^{\left( 2\right) }\theta _{j}$. The available \emph{relevant
information} can be written in the form of the following $2n$ information
constraint equations,%
\begin{equation}
\left\langle x_{j}\right\rangle =\dint\limits_{-\infty }^{+\infty
}dx_{j}x_{j}P_{j}\left( x_{j}|^{\left( 1\right) }\theta _{j}\text{,}^{\left(
2\right) }\theta _{j}\right) \text{, }\Delta x_{j}=\left[ \dint\limits_{-%
\infty }^{+\infty }dx_{j}\left( x_{j}-\left\langle x_{j}\right\rangle
\right) ^{2}P_{j}\left( x_{j}|^{\left( 1\right) }\theta _{j}\text{,}^{\left(
2\right) }\theta _{j}\right) \right] ^{\frac{1}{2}}\text{.}  \label{C1}
\end{equation}%
The probability distributions $P_{j}$ in (\ref{C1}) are constrained by the
conditions of normalization,%
\begin{equation}
\dint\limits_{-\infty }^{+\infty }dx_{j}P_{j}\left( x_{j}|^{\left( 1\right)
}\theta _{j}\text{,}^{\left( 2\right) }\theta _{j}\right) =1\text{.}
\label{C2}
\end{equation}%
Information theory identifies the Gaussian distribution as the maximum
entropy distribution if only the expectation value and the variance are
known \cite{tribus}. Maximum relative Entropy methods \cite{caticha(REII),
caticha-giffin, caticha2} allow us to associate a probability distribution $%
P\left( X|\Theta \right) $ to each point in the space of states $\Theta $.
The distribution that best reflects the information contained in the prior
distribution $m\left( X\right) $ updated by the information $\left(
\left\langle x_{j}\right\rangle \text{, }\Delta x_{j}\right) $ is obtained
by maximizing the relative entropy, 
\begin{equation}
S\left( \Theta \right) =-\int d^{l}XP\left( X|\Theta \right) \log \left( 
\frac{P\left( X|\Theta \right) }{m\left( X\right) }\right) \text{,}
\label{RE}
\end{equation}%
where $m\left( X\right) $ is the prior probability distribution. As a
working hypothesis, the prior $m\left( X\right) $ is set to be uniform since
we assume the lack of prior available information about the system \cite%
{jay57}. We assume uncoupled constraints among microvariables\textbf{\ }$%
x_{j}$\textbf{.} In other words, we assume that information about
correlations between the microvariables need not to be tracked. Therefore,
upon maximizing (\ref{RE}) given the constraints (\ref{C1}) and (\ref{C2}),
we obtain%
\begin{equation}
P\left( X|\Theta \right) =\dprod\limits_{j=1}^{n}P_{j}\left( x_{j}|^{\left(
1\right) }\theta _{j}\text{,}^{\left( 2\right) }\theta _{j}\right) 
\label{PDG}
\end{equation}%
where%
\begin{equation}
P_{j}\left( x_{j}|^{\left( 1\right) }\theta _{j}\text{,}^{\left( 2\right)
}\theta _{j}\right) =\left( 2\pi \sigma _{j}^{2}\right) ^{-\frac{1}{2}}\exp %
\left[ -\frac{\left( x_{j}-\mu _{j}\right) ^{2}}{2\sigma _{j}^{2}}\right] 
\end{equation}%
and,\textbf{\ }in standard notation for Gaussians, $^{\left( 1\right)
}\theta _{^{j}}\overset{\text{def}}{=}\left\langle x_{j}\right\rangle \equiv
\mu _{j}$, $^{\left( 2\right) }\theta _{j}\overset{\text{def}}{=}\Delta
x_{j}\equiv \sigma _{j}$. The probability distribution (\ref{PDG}) encodes
the available information concerning the system. The statistical manifold $%
\mathcal{M}_{S}$ associated to (\ref{PDG}) is formally defined as follows,%
\begin{equation}
\mathcal{M}_{S}=\left\{ P\left( X|\Theta \right) =\underset{j=1}{\overset{n}{%
\dprod }}P_{j}\left( x_{j}|\mu _{j}\text{, }\sigma _{j}\right) \right\} 
\text{,}  \label{manifold}
\end{equation}%
where $X\in 
\mathbb{R}
^{n}$ and $\Theta $ belongs to the $2n$-dimensional parameter space $%
\mathcal{D}_{\Theta }=\left[ \mathcal{I}_{\mu }\times \mathcal{I}_{\sigma }%
\right] ^{n}$\textbf{. }The parameter space $\mathcal{D}_{\Theta }$
(homeomorphic to\textbf{\ }$\mathcal{M}_{S}$\textbf{)} is the direct product
of the parameter subspaces $\mathcal{I}_{\mu }$ and $\mathcal{I}_{\sigma }$%
\textbf{, }where (in the Gaussian case, unless specified otherwise) $%
\mathcal{I}_{\mu }=\left( -\infty \text{, }+\infty \right) _{\mu }$ and $%
\mathcal{I}_{\sigma }=\left( 0\text{, }+\infty \right) _{\sigma }$\textbf{. }%
The line element $ds^{2}=g_{\mu \nu }\left( \Theta \right) d\Theta ^{\mu
}d\Theta ^{\nu }$ arising from (\ref{PDG}) is \cite{cafaroIJTP},%
\begin{equation}
ds_{\mathcal{M}_{s}}^{2}\overset{\text{def}}{=}\dsum\limits_{j=1}^{n}\left( 
\frac{1}{\sigma _{j}^{2}}d\mu _{j}^{2}+\frac{2}{\sigma _{j}^{2}}d\sigma
_{j}^{2}\right) \text{, }
\end{equation}%
with $\mu $, $\nu =1$,..., $2n$.

\subsection{Gaussian Statistical Models in Presence of Microcorrelations}

Coupled constraints would lead to a "generalized" product rule and to a
metric tensor with non-trivial off-diagonal elements (covariance terms). In
presence of correlated degrees of freedom $\left\{ x_{j}\right\} $, the
"generalized" product rule becomes,%
\begin{equation}
P_{\text{tot}}\left( x_{1}\text{,..., }x_{n}\right)
=\dprod\limits_{j=1}^{n}P_{j}\left( x_{j}\right) \overset{\text{correlations}%
}{\longrightarrow }P_{\text{tot}}^{\prime }\left( x_{1}\text{,..., }%
x_{n}\right) \neq \dprod\limits_{j=1}^{n}P_{j}\left( x_{j}\right) \text{,}
\end{equation}%
where,%
\begin{equation}
P_{\text{tot}}^{\prime }\left( x_{1}\text{,..., }x_{n}\right) \overset{\text{%
def}}{=}P_{n}\left( x_{n}|x_{1}\text{,..., }x_{n-1}\right) P_{n-1}\left(
x_{n-1}|x_{1}\text{,..., }x_{n-2}\right) \text{...}P_{2}\left(
x_{2}|x_{1}\right) P_{1}\left( x_{1}\right) \text{.}
\end{equation}%
\textbf{Correlations} among the degrees of freedom may be introduced in
terms of the following information-constraints,%
\begin{equation}
x_{j}=f_{j}\left( x_{1}\text{,..., }x_{j-1}\right) \text{, }\forall j=2\text{%
,..., }n\text{.}
\end{equation}%
In such a case, we obtain%
\begin{equation}
P_{\text{tot}}^{\prime }\left( x_{1}\text{,..., }x_{n}\right) =\delta \left(
x_{n}-f_{n}\left( x_{1}\text{,..., }x_{n-1}\right) \right) \delta \left(
x_{n-1}-f_{n-1}\left( x_{1}\text{,..., }x_{n-2}\right) \right) ...\delta
\left( x_{2}-f_{2}\left( x_{1}\right) \right) P_{1}\left( x_{1}\right) \text{%
,}
\end{equation}%
where the $j$-th probability distribution $P_{j}\left( x_{j}\right) $ is
given by,%
\begin{equation}
P_{j}\left( x_{j}\right) =\int \text{...}\int dx_{1}\text{...}%
dx_{j-1}dx_{j+1}\text{...}dx_{n}P_{\text{tot}}^{\prime }\left( x_{1}\text{%
,..., }x_{n}\right) \text{.}
\end{equation}%
\textbf{A} formal manner in which correlations are introduced in probability
theory is as follows. Given two arbitrary randomly distributed variables $%
x_{1}$\textbf{\ }and $x_{2}$\textbf{,} consider the problem of finding a
linear expression of the form $\tilde{c}_{1}+\tilde{c}_{2}x_{2}$\textbf{,}
involving \emph{real} constants $\tilde{c}_{1}$ and $\tilde{c}_{2}$ such
that $\tilde{c}_{1}+\tilde{c}_{2}x_{2}$ is the best "mean square
approximation" to $x_{1}$. The best approximation is such that%
\begin{equation}
\left\langle \left( x_{1}-\tilde{c}_{1}-\tilde{c}_{2}x_{2}\right)
^{2}\right\rangle =\min_{c_{1}\text{, }c_{2}}\left\langle \left(
x_{1}-c_{1}-c_{2}x_{2}\right) ^{2}\right\rangle \text{,}  \label{pr}
\end{equation}%
where the minimum is taken with respect to all \emph{real} constants $c_{1}$
and $c_{2}$. To solve this problem, let%
\begin{equation}
\mu _{1}=\left\langle x_{1}\right\rangle \text{, }\sigma
_{1}^{2}=\left\langle \left( x_{1}-\left\langle x_{1}\right\rangle \right)
^{2}\right\rangle \text{, }\mu _{2}=\left\langle x_{2}\right\rangle \text{, }%
\sigma _{2}^{2}=\left\langle \left( x_{2}-\left\langle x_{2}\right\rangle
\right) ^{2}\right\rangle
\end{equation}%
and introduce the quantity \cite{roz},%
\begin{equation}
r\overset{\text{def}}{=}\frac{\left\langle \left( x_{1}-\left\langle
x_{1}\right\rangle \right) \left( x_{2}-\left\langle x_{2}\right\rangle
\right) \right\rangle }{\sigma _{1}\sigma _{2}}=\frac{\left\langle
x_{1}x_{2}\right\rangle -\mu _{1}\mu _{2}}{\sigma _{1}\sigma _{2}}\text{.}
\end{equation}%
The quantity\textbf{\ }$r$\textbf{\ }is the so-called correlation
coefficient of the random variables $x_{1}$and $x_{2}$. For the sake of
convenience, we may introduce the "normalized" random variables,%
\begin{equation}
\eta _{1}\overset{\text{def}}{=}\frac{x_{1}-\mu _{1}}{\sigma _{1}}\text{
and, }\eta _{2}\overset{\text{def}}{=}\frac{x_{2}-\mu _{2}}{\sigma _{2}}%
\text{.}
\end{equation}%
The problem in (\ref{pr})\ can now be reduced to,%
\begin{equation}
\min_{c_{1}\text{, }c_{2}}\left\langle \left( \eta _{1}-c_{1}-c_{2}\eta
_{2}\right) ^{2}\right\rangle =\min_{c_{1}\text{, }c_{2}}\left[ \left(
1-r^{2}\right) +c_{1}^{2}+\left( r-c_{2}\right) ^{2}\right] =1-r^{2}\geq 0%
\text{.}
\end{equation}%
The minimum is achieved for $c_{1}=0$ and $c_{2}=r$, where $r$ lies in the
interval $-1\leq r\leq +1$.

In our work, correlations among the microscopic degrees of freedom of the
system $\left\{ x_{j}\right\} $ (microcorrelations) are conventionally
introduced by means of the correlation coefficients $r_{ij}^{\left( \text{%
micro}\right) }$, 
\begin{equation}
r_{ij}^{\left( \text{micro}\right) }=r\left( x_{i}\text{, }x_{j}\right) 
\overset{\text{def}}{=}\frac{\left\langle x_{i}x_{j}\right\rangle
-\left\langle x_{i}\right\rangle \left\langle x_{j}\right\rangle }{\sigma
_{i}\sigma _{j}}\text{, with }\sigma _{i}=\sqrt{\left\langle \left(
x_{i}-\left\langle x_{i}\right\rangle \right) ^{2}\right\rangle }\text{,}
\end{equation}%
with $r_{ij}^{\left( \text{micro}\right) }\in \left( -1\text{, }1\right) $
and $i$, $j=1$,..., $n$. For the $2n$-dimensional Gaussian statistical model
in presence of microcorrelations, the system is described by the following
probability distribution $P\left( X|\Theta \right) $,%
\begin{equation}
P\left( X|\Theta \right) =\frac{1}{\left[ \left( 2\pi \right) ^{n}\det
C\left( \Theta \right) \right] ^{\frac{1}{2}}}\exp \left[ -\frac{1}{2}\left(
X-M\right) ^{t}\cdot C^{-1}\left( \Theta \right) \cdot \left( X-M\right) %
\right] \neq \dprod\limits_{j=1}^{n}\left( 2\pi \sigma _{j}^{2}\right) ^{-%
\frac{1}{2}}\exp \left[ -\frac{\left( x_{j}-\mu _{j}\right) ^{2}}{2\sigma
_{j}^{2}}\right] \text{,}  \label{CG}
\end{equation}%
where $X=\left( x_{1}\text{,..., }x_{n}\right) $, $M=\left( \mu _{1}\text{%
,..., }\mu _{n}\right) $ and $C\left( \Theta \right) $ is the $\left(
2n\times 2n\right) $-dimensional (non-singular) covariance matrix.

\section{The Model}

In this Section we focus on microcorrelated Gaussian statistical models with 
$2n=4$. For $n=2$, (\ref{CG}) leads to the probability distribution $P\left(
x\text{, }y|\mu _{x}\text{, }\sigma _{x}\text{, }\mu _{y}\text{, }\sigma
_{y}\right) $ which takes the form,%
\begin{equation}
P\left( x\text{, }y|\mu _{x}\text{, }\sigma _{x}\text{, }\mu _{y}\text{, }%
\sigma _{y}\right) =\frac{\exp \left\{ -\frac{1}{2\left( 1-r^{2}\right) }%
\left[ \frac{\left( x-\mu _{x}\right) ^{2}}{\sigma _{x}^{2}}-2r\frac{\left(
x-\mu _{x}\right) \left( y-\mu _{y}\right) }{\sigma _{x}\sigma _{y}}+\frac{%
\left( y-\mu _{y}\right) ^{2}}{\sigma _{y}^{2}}\right] \right\} }{2\pi
\sigma _{x}\sigma _{y}\sqrt{1-r^{2}}}\text{,}  \label{2g}
\end{equation}%
where $\sigma _{x}>0$, $\sigma _{y}>0$, $r\in \left( -1\text{, }+1\right) $.
Substituting (\ref{2g}) in (\ref{FRM}), the Fisher-Rao information metric $%
g_{\mu \nu }\left( \mu _{x}\text{, }\sigma _{x}\text{, }\mu _{y}\text{, }%
\sigma _{y}\text{; }r\right) $ becomes,%
\begin{equation}
g_{\mu \nu }\left( \mu _{x}\text{, }\sigma _{x}\text{, }\mu _{y}\text{, }%
\sigma _{y}\text{; }r\right) =\left( 
\begin{array}{cccc}
-\frac{1}{\sigma _{x}^{2}\left( r^{2}-1\right) } & 0 & \frac{r}{\sigma
_{x}\sigma _{y}\left( r^{2}-1\right) } & 0 \\ 
0 & -\frac{2-r^{2}}{\sigma _{x}^{2}\left( r^{2}-1\right) } & 0 & \frac{r^{2}%
}{\sigma _{x}\sigma _{y}\left( r^{2}-1\right) } \\ 
\frac{r}{\sigma _{x}\sigma _{y}\left( r^{2}-1\right) } & 0 & \frac{1}{\sigma
_{y}^{2}\left( r^{2}-1\right) } & 0 \\ 
0 & \frac{r^{2}}{\sigma _{x}\sigma _{y}\left( r^{2}-1\right) } & 0 & -\frac{%
2-r^{2}}{\sigma _{y}^{2}\left( r^{2}-1\right) }%
\end{array}%
\right) \text{.}  \label{corr}
\end{equation}%
The infinitesimal line element $ds_{\mathcal{M}_{S}}^{2}$ relative to $%
g_{\mu \nu }\left( \mu _{x}\text{, }\sigma _{x}\text{, }\mu _{y}\text{, }%
\sigma _{y}\text{; }r\right) $ is given by,%
\begin{eqnarray}
ds_{\mathcal{M}_{S}}^{2} &=&g_{11}\left( \sigma _{x}\text{; }r\right) d\mu
_{x}^{2}+g_{33}\left( \sigma _{y}\text{; }r\right) d\mu
_{y}^{2}+g_{22}\left( \sigma _{x}\text{; }r\right) d\sigma
_{x}^{2}+g_{44}\left( \sigma _{y}\text{; }r\right) d\sigma
_{y}^{2}+2g_{13}\left( \sigma _{x}\text{, }\sigma _{y}\text{; }r\right) d\mu
_{x}d\mu _{y}  \notag \\
&&  \notag \\
&&+2g_{24}\left( \sigma _{x}\text{, }\sigma _{y}\text{; }r\right) d\sigma
_{x}d\sigma _{y}\text{,}  \label{le1}
\end{eqnarray}%
where,%
\begin{eqnarray}
g_{11}\left( \sigma _{x}\text{; }r\right) &=&-\frac{1}{\sigma _{x}^{2}\left(
r^{2}-1\right) }\text{, }g_{13}\left( \sigma _{x}\text{, }\sigma _{y}\text{; 
}r\right) =\frac{r}{\sigma _{x}\sigma _{y}\left( r^{2}-1\right) }\text{, }%
g_{22}\left( \sigma _{x}\text{; }r\right) =-\frac{2-r^{2}}{\sigma
_{x}^{2}\left( r^{2}-1\right) }\text{,}  \notag \\
&&  \notag \\
g_{24}\left( \sigma _{x}\text{, }\sigma _{y}\text{; }r\right) &=&\frac{r^{2}%
}{\sigma _{x}\sigma _{y}\left( r^{2}-1\right) }\text{, }g_{31}\left( \sigma
_{x}\text{, }\sigma _{y}\text{; }r\right) =\frac{r}{\sigma _{x}\sigma
_{y}\left( r^{2}-1\right) }\text{, }g_{33}\left( \sigma _{y}\text{; }%
r\right) =-\frac{1}{\sigma _{y}^{2}\left( r^{2}-1\right) }\text{,}  \notag \\
&&  \notag \\
g_{42}\left( \sigma _{x}\text{, }\sigma _{y}\text{; }r\right) &=&\frac{r^{2}%
}{\sigma _{x}\sigma _{y}\left( r^{2}-1\right) }\text{, }g_{44}\left( \sigma
_{y}\text{; }r\right) =-\frac{2-r^{2}}{\sigma _{y}^{2}\left( r^{2}-1\right) }%
\text{.}
\end{eqnarray}%
The analytical study of the IGAC arising on a curved statistical manifold
with infinitesimal line element given by $ds_{\mathcal{M}_{S}}^{2}$ in (\ref%
{le1}) turns out to be rather difficult. Hence, as working hypothesis, we
are going to assume two correlated Gaussian-distributed microvariables
characterized by the same variance, that is we assume $\sigma _{x}=\sigma
_{y}\equiv \sigma $. Thus, the simplified line element becomes,%
\begin{equation}
ds_{\mathcal{M}_{S}}^{2}=g_{11}\left( \sigma _{x}\text{; }r\right) d\mu
_{x}^{2}+g_{33}\left( \sigma _{y}\text{; }r\right) d\mu
_{y}^{2}+2g_{13}\left( \sigma \text{; }r\right) d\mu _{x}d\mu _{y}+\left[
g_{22}\left( \sigma \text{; }r\right) +g_{44}\left( \sigma \text{; }r\right)
+2g_{24}\left( \sigma \text{; }r\right) \right] d\sigma ^{2}\text{.}
\label{le}
\end{equation}%
The new Fisher-Rao matrix $g_{\mu \nu }\left( \mu _{x}\text{, }\mu _{y}\text{%
, }\sigma \text{; }r\right) $ associated with line element $ds_{\mathcal{M}%
_{S}}^{2}$ in (\ref{le}) becomes,%
\begin{equation}
g_{\mu \nu }\left( \mu _{x}\text{, }\mu _{y}\text{, }\sigma \text{; }%
r\right) =\frac{1}{\sigma ^{2}}\left( 
\begin{array}{ccc}
-\frac{1}{r^{2}-1} & \frac{r}{2\left( r^{2}-1\right) } & 0 \\ 
\frac{r}{2\left( r^{2}-1\right) } & -\frac{1}{r^{2}-1} & 0 \\ 
0 & 0 & 4%
\end{array}%
\right) \text{.}  \label{cim}
\end{equation}%
We will study the information dynamics on curved statistical manifolds $%
\mathcal{M}_{S}^{\left( \text{correlations}\right) }$ and $\mathcal{M}%
_{S}^{\left( \text{no-correlations}\right) }$ with\textbf{\ }infinitesimal
line elements $\left( ds_{\mathcal{M}_{S}}^{2}\right) ^{\text{correlations}}$
and $\left( ds_{\mathcal{M}_{S}}^{2}\right) ^{\text{no-correlations}}$,
respectively. The line element $\left( ds_{\mathcal{M}_{S}}^{2}\right) ^{%
\text{correlations}}$ is defined by,%
\begin{equation}
\left( ds_{\mathcal{M}_{S}}^{2}\right) ^{\text{correlations}}\overset{\text{%
def}}{=}\frac{1}{\sigma ^{2}}\left( \frac{1}{1-r^{2}}d\mu _{x}^{2}+\frac{1}{%
1-r^{2}}d\mu _{y}^{2}-\frac{2r}{1-r^{2}}d\mu _{x}d\mu _{y}+4d\sigma
^{2}\right) \text{,}  \label{cfr}
\end{equation}%
while $\left( ds_{\mathcal{M}_{S}}^{2}\right) ^{\text{no-correlations}}$ is
obtained from $\left( ds_{\mathcal{M}_{S}}^{2}\right) ^{\text{correlations}}$
in the limit that $r$ approaches zero\textbf{.}

\subsection{Information Geometry of The Model}

Consider the \ information dynamics of the Model introduced in Section II.
The Fisher-Rao line element $\left( ds_{\mathcal{M}_{S}}^{2}\right) ^{\text{%
correlations}}$ of such statistical model $\mathcal{M}_{S}^{\left( \text{%
correlations}\right) }$ is given in (\ref{cfr}). The inverse metric tensor $%
g^{\mu \nu }\left( \mu _{x}\text{, }\mu _{y}\text{, }\sigma \text{; }%
r\right) $ is given by,%
\begin{equation}
g^{\mu \nu }\left( \mu _{x}\text{, }\mu _{y}\text{, }\sigma \text{; }%
r\right) =\sigma ^{2}\left( 
\begin{array}{ccc}
\frac{4\left( r^{2}-1\right) }{r^{2}-4} & \frac{2r\left( r^{2}-1\right) }{%
r^{2}-4} & 0 \\ 
\frac{2r\left( r^{2}-1\right) }{r^{2}-4} & \frac{4\left( r^{2}-1\right) }{%
r^{2}-4} & 0 \\ 
0 & 0 & \frac{1}{4}%
\end{array}%
\right) \text{.}  \label{IM}
\end{equation}%
The metric tensor $g_{\mu \nu }\left( \mu _{x}\text{, }\mu _{y}\text{, }%
\sigma \text{; }r\right) $ and its inverse $g^{\mu \nu }\left( \mu _{x}\text{%
, }\mu _{y}\text{, }\sigma \text{; }r\right) $ are necessary to determine
the Christoffel connection coefficients $\Gamma _{ij}^{k}$ of the manifold $%
\mathcal{M}_{S}^{\left( \text{correlations}\right) }$. Recall that the
connection coefficients $\Gamma _{ij}^{k}$ are defined as \cite{felice},%
\begin{equation}
\Gamma _{ij}^{k}\overset{\text{def}}{=}\frac{1}{2}g^{km}\left( \partial
_{i}g_{mj}+\partial _{j}g_{im}-\partial _{m}g_{ij}\right) \text{.}  \label{c}
\end{equation}%
In our case, the non-vanishing connection coefficients are given by,%
\begin{eqnarray}
\Gamma _{11}^{3} &=&-\frac{1}{4}\frac{1}{r^{2}-1}\frac{1}{\sigma }\text{, }%
\Gamma _{12}^{3}=\Gamma _{21}^{3}=\frac{r}{8\left( r^{2}-1\right) }\frac{1}{%
\sigma }\text{, }\Gamma _{13}^{1}=\Gamma _{31}^{1}=-\frac{1}{\sigma }\text{, 
}\Gamma _{22}^{3}=-\frac{1}{4}\frac{1}{r^{2}-1}\frac{1}{\sigma }\text{,} 
\notag \\
&&  \notag \\
\text{ }\Gamma _{23}^{2} &=&\Gamma _{32}^{2}=-\frac{1}{\sigma }\text{, }%
\Gamma _{33}^{3}=-\frac{1}{\sigma }\text{.}  \label{CS}
\end{eqnarray}%
Once the non-vanishing components of $\Gamma _{ij}^{k}$ are obtained, we
compute the Ricci curvature tensor $\mathcal{R}_{ij}$ defined as \cite%
{felice},%
\begin{equation}
\mathcal{R}_{ij}\overset{\text{def}}{=}\partial _{k}\Gamma
_{ij}^{k}-\partial _{j}\Gamma _{ik}^{k}+\Gamma _{ij}^{k}\Gamma
_{kn}^{n}-\Gamma _{ik}^{m}\Gamma _{jm}^{k}\text{.}  \label{RICCI}
\end{equation}%
Substituting (\ref{CS}) in (\ref{RICCI}), we obtain the non-vanishing Ricci
curvature tensor components $\mathcal{R}_{ij}$,%
\begin{equation}
R_{11}=\frac{1}{2\left( r^{2}-1\right) }\frac{1}{\sigma ^{2}}\text{, }%
R_{12}=R_{21}=-\frac{r}{4\left( r^{2}-1\right) }\frac{1}{\sigma ^{2}}\text{, 
}R_{22}=\frac{1}{2\left( r^{2}-1\right) }\frac{1}{\sigma ^{2}}\text{, }%
R_{33}=-\frac{2}{\sigma ^{2}}\text{.}  \label{R}
\end{equation}%
Finally, we compute Ricci scalar curvature $\mathcal{R}_{\mathcal{M}%
_{s}}\left( r\right) $,%
\begin{equation}
\mathcal{R}_{\mathcal{M}_{s}}\left( r\right) \overset{\text{def}}{=}\mathcal{%
R}_{ij}g^{ij}\text{.}  \label{scalar}
\end{equation}%
Substituting (\ref{R}) and (\ref{IM}) in (\ref{scalar})\textbf{,} $\mathcal{R%
}_{\mathcal{M}_{s}}\left( r\right) $ becomes\textbf{,}%
\begin{equation}
\mathcal{R}_{\mathcal{M}_{s}}\left( r\right)
=g^{11}R_{11}+2g^{12}R_{12}+g^{22}R_{22}+g^{33}R_{33}=-\frac{3}{2}\text{.}
\end{equation}%
Therefore, we conclude that $\mathcal{M}_{s}^{\left( \text{correlations}%
\right) }$\textbf{\ }is a curved statistical manifold of constant negative
curvature.

\subsection{Information Dynamics on $\mathcal{M}_{s}$}

The information dynamics can be derived from a standard principle of least
action of Jacobi type \cite{caticha1}. The geodesic equations for the
macrovariables of the Gaussian ED model are given by\textit{\ nonlinear}
second order coupled ordinary differential equations,%
\begin{equation}
\frac{d^{2}\Theta ^{\mu }}{d\tau ^{2}}+\Gamma _{\nu \rho }^{\mu }\frac{%
d\Theta ^{\nu }}{d\tau }\frac{d\Theta ^{\rho }}{d\tau }=0\text{.}  \label{GE}
\end{equation}%
The geodesic equations in (\ref{GE}) describe a \textit{reversible} dynamics
whose solution is the trajectory between an initial $\Theta ^{\left( \text{%
initial}\right) }$ and a final macrostate $\Theta ^{\left( \text{final}%
\right) }$. The trajectory can be equally well traversed in both directions.
In the case under consideration, substituting (\ref{cim}) in (\ref{GE}), the
three geodesic equations become,%
\begin{eqnarray}
0 &=&\frac{d^{2}\mu _{x}\left( \tau \right) }{d\tau ^{2}}-\frac{2}{\sigma
\left( \tau \right) }\frac{d\mu _{x}\left( \tau \right) }{d\tau }\frac{%
d\sigma \left( \tau \right) }{d\tau }\text{,}  \notag \\
&&  \notag \\
\text{ }0 &=&\frac{d^{2}\mu _{y}\left( \tau \right) }{d\tau ^{2}}-\frac{2}{%
\sigma \left( \tau \right) }\frac{d\mu _{y}\left( \tau \right) }{d\tau }%
\frac{d\sigma \left( \tau \right) }{d\tau }\text{,}  \notag \\
&&  \notag \\
0 &=&\frac{d^{2}\sigma \left( \tau \right) }{d\tau ^{2}}-\frac{1}{\sigma
\left( \tau \right) }\left( \frac{d\sigma \left( \tau \right) }{d\tau }%
\right) ^{2}-\frac{1}{4}\frac{1}{r^{2}-1}\frac{1}{\sigma \left( \tau \right) 
}\left( \frac{d\mu _{x}\left( \tau \right) }{d\tau }\right) ^{2}-\frac{1}{4}%
\frac{1}{r^{2}-1}\frac{1}{\sigma \left( \tau \right) }\left( \frac{d\mu
_{y}\left( \tau \right) }{d\tau }\right) ^{2}+  \notag \\
&&  \notag \\
&&+\frac{r}{4\left( r^{2}-1\right) }\frac{1}{\sigma }\frac{d\mu _{x}\left(
\tau \right) }{d\tau }\frac{d\mu _{y}\left( \tau \right) }{d\tau }\text{.}
\label{GEE}
\end{eqnarray}%
Integration of the above coupled system of differential equations is non
trivial. A detailed derivation of the geodesic paths is given in the
Appendix. After integration of (\ref{GEE}), the geodesic trajectories become,%
\begin{eqnarray}
\mu _{x}\left( \tau \text{; }r\right) &=&-\frac{2\sigma _{0}A_{1}}{\sqrt{%
\mathcal{A}\left( r\right) }}\frac{1}{1+\exp \left( 2\sigma _{0}\sqrt{%
\mathcal{A}\left( r\right) }\tau \right) }\text{, }\mu _{y}\left( \tau \text{%
; }r\right) =-\frac{2\sigma _{0}A_{2}}{\sqrt{\mathcal{A}\left( r\right) }}%
\frac{1}{1+\exp \left( 2\sigma _{0}\sqrt{\mathcal{A}\left( r\right) }\tau
\right) }\text{, }  \notag \\
&&  \notag \\
\sigma \left( \tau \text{; }r\right) &=&2\sigma _{0}\frac{\exp \left( \sigma
_{0}\sqrt{\mathcal{A}\left( r\right) }\tau \right) }{1+\exp \left( 2\sigma
_{0}\sqrt{\mathcal{A}\left( r\right) }\tau \right) }\text{,}  \label{gsol1}
\end{eqnarray}%
where,%
\begin{equation}
\mathcal{A}\left( r\right) \overset{\text{def}}{=}\frac{%
A_{1}^{2}+A_{2}^{2}-rA_{1}A_{2}}{4\left( 1-r^{2}\right) }\text{.}
\end{equation}%
Notice that for any \emph{real} value of $A_{1}$ and $A_{2}$, $0\leq \left(
A_{1}-A_{2}\right) ^{2}=A_{1}^{2}+A_{2}^{2}-2A_{1}A_{2}\leq
A_{1}^{2}+A_{2}^{2}-rA_{1}A_{2}$ and $4\left( 1-r^{2}\right) \geq 0$ for $%
r\in \left( -1\text{, }1\right) $.\ It then follows that, $\mathcal{A}\left(
r\right) \geq 0$. Note that $\sigma \left( \tau \text{; }r\right) \in \left(
0\text{, }+\infty \right) $ while $\mu _{x}\left( \tau \text{; }r\right) $
and $\mu _{y}\left( \tau \text{; }r\right) \in \left( -\infty \text{, }%
+\infty \right) $.

\section{The Information Geometric Complexity and Microcorrelations}

We recall that a suitable indicator of temporal complexity within the IGAC
framework is provided by the \emph{information geometric entropy} (IGE) $%
\mathcal{S}_{\mathcal{M}_{s}}\left( \tau \right) $ \cite{carlo-tesi,
carlo-CSF},%
\begin{equation}
\mathcal{S}_{\mathcal{M}_{s}}\left( \tau \right) \overset{\text{def}}{=}\log 
\widetilde{\emph{vol}}\left[ \mathcal{D}_{\Theta }^{\left( \text{geodesic}%
\right) }\left( \tau \right) \right] \text{.}
\end{equation}%
The \emph{information geometric complexity} (IGC) is defined as the average
dynamical statistical volume $\widetilde{\emph{vol}}\left[ \mathcal{D}%
_{\Theta }^{\left( \text{geodesic}\right) }\left( \tau \right) \right] $
given by,%
\begin{equation}
\widetilde{\emph{vol}}\left[ \mathcal{D}_{\Theta }^{\left( \text{geodesic}%
\right) }\left( \tau \right) \right] \overset{\text{def}}{=}\lim_{\tau
\rightarrow \infty }\left( \frac{1}{\tau }\int_{0}^{\tau }d\tau ^{\prime }%
\emph{vol}\left[ \mathcal{D}_{\Theta }^{\left( \text{geodesic}\right)
}\left( \tau ^{\prime }\right) \right] \right) \text{,}
\end{equation}%
where the "tilde" symbol denotes the operation of temporal average. The
volume $\emph{vol}\left[ \mathcal{D}_{\Theta }^{\left( \text{geodesic}%
\right) }\left( \tau ^{\prime }\right) \right] $ is given by,%
\begin{equation}
\emph{vol}\left[ \mathcal{D}_{\Theta }^{\left( \text{geodesic}\right)
}\left( \tau ^{\prime }\right) \right] \overset{\text{def}}{=}\int_{\mathcal{%
D}_{\Theta }^{\left( \text{geodesic}\right) }\left( \tau ^{\prime }\right)
}\rho _{\left( \mathcal{M}_{s}\text{, }g\right) }\left( \theta ^{1}\text{%
,..., }\theta ^{N}\right) d^{N}\Theta \text{,}  \label{v}
\end{equation}%
where $N$ is the dimensionality of the statistical manifold $\mathcal{M}_{s}$
and $\rho _{\left( \mathcal{M}_{s}\text{, }g\right) }\left( \theta ^{1}\text{%
,..., }\theta ^{N}\right) $ is the so-called Fisher density and equals the
square root of the determinant of the metric tensor $g_{\mu \nu }\left(
\Theta \right) $,%
\begin{equation}
\rho _{\left( \mathcal{M}_{s}\text{, }g\right) }\left( \theta ^{1}\text{%
,..., }\theta ^{N}\right) \overset{\text{def}}{=}\sqrt{g\left( \theta ^{1}%
\text{,..., }\theta ^{N}\right) }\text{.}
\end{equation}%
The integration space $\mathcal{D}_{\Theta }^{\left( \text{geodesic}\right)
}\left( \tau ^{\prime }\right) $ in (\ref{v}) is defined as follows,%
\begin{equation}
\mathcal{D}_{\Theta }^{\left( \text{geodesic}\right) }\left( \tau ^{\prime
}\right) \overset{\text{def}}{=}\left\{ \Theta \equiv \left( \theta ^{1}%
\text{,..., }\theta ^{N}\right) :\theta ^{k}\left( 0\right) \leq \theta
^{k}\leq \theta ^{k}\left( \tau ^{\prime }\right) \right\} \text{,}
\label{is}
\end{equation}%
where $k=1$,.., $N$ and $\theta ^{k}\equiv \theta ^{k}\left( s\right) $ with 
$0\leq s\leq \tau ^{\prime }$ such that $\theta ^{k}\left( s\right) $
satisfies (\ref{GE}). The integration space $\mathcal{D}_{\Theta }^{\left( 
\text{geodesic}\right) }\left( \tau ^{\prime }\right) $ in (\ref{is}) is an $%
N$-dimensional subspace of the whole (permitted) parameter space $\mathcal{D}%
_{\Theta }^{\left( \text{tot}\right) }$. The elements of $\mathcal{D}%
_{\Theta }^{\left( \text{geodesic}\right) }\left( \tau ^{\prime }\right) $
are the $N$-dimensional macrovariables $\left\{ \Theta \right\} $ whose
components $\theta ^{k}$ are bounded by specified limits of integration $%
\theta ^{k}\left( 0\right) $ and $\theta ^{k}\left( \tau ^{\prime }\right) $
with $k=1$,.., $N$. The limits of integration are obtained via integration
of the geodesic equations. Formally, the IGE $\mathcal{S}_{\mathcal{M}%
_{s}}\left( \tau \right) $ is defined in terms of a averaged parametric $%
\left( N+1\right) $-fold integral ($\tau $ is the parameter) over the
multidimensional geodesic paths connecting $\Theta \left( 0\right) $ to $%
\Theta \left( \tau \right) $. The quantity $\emph{vol}\left[ \mathcal{D}%
_{\Theta }^{\left( \text{geodesic}\right) }\left( \tau ^{\prime }\right) %
\right] $ is the volume of the effective parameter space explored by the
system at time $\tau ^{\prime }$. The temporal average has been introduced
in order to average out the possibly very complex fine details of the
entropic dynamical description of the system on $\mathcal{M}_{S}$. Thus, we
provide a coarse-grained-like (or randomized-like) inferential description
of the system's chaotic dynamics. The long-term asymptotic temporal behavior
is adopted in order to properly characterize dynamical indicators of
chaoticity (for instance,\ Lyapunov exponents and the Kolmogorov-Sinai
dynamical entropy) eliminating the effects of transient effects which enter
the computation of the expected value of $\emph{vol}\left[ \mathcal{D}%
_{\Theta }^{\left( \text{geodesic}\right) }\left( \tau ^{\prime }\right) %
\right] $\textbf{. }In chaotic transients, one observes that typical initial
conditions behave in an apparently chaotic manner for a possibly long time,
but then asymptotically approach a nonchaotic attractor in a rapid fashion.

In the case under consideration, $\emph{vol}\left[ \mathcal{D}_{\Theta
}^{\left( \text{geodesic}\right) }\left( \tau \right) \right] _{\text{cor.}}$
is given by,%
\begin{equation}
\emph{vol}\left[ \mathcal{D}_{\Theta }^{\left( \text{geodesic}\right)
}\left( \tau \right) \right] _{\text{cor.}}=\int \sqrt{g}d\mu _{x}d\mu
_{y}d\sigma =-\frac{A_{1}A_{2}}{2\mathcal{A}\left( r\right) }\sqrt{g^{\prime
}}\exp \left( -\sigma _{0}\sqrt{\mathcal{A}\left( r\right) }\tau \right) 
\text{,}  \label{ve}
\end{equation}%
with%
\begin{equation}
g\left( r\right) \overset{\text{def}}{=}\frac{4\left( 4-r^{2}\right) }{%
\left( 2-2r^{2}\right) ^{2}}\frac{1}{\sigma ^{6}}\text{ and }g^{\prime \frac{%
1}{2}}\left( r\right) \overset{\text{def}}{=}\sqrt{\frac{4\left(
4-r^{2}\right) }{\left( 2-2r^{2}\right) ^{2}}}\text{.}
\end{equation}%
Since $\emph{vol}\left[ \mathcal{D}_{\Theta }^{\left( \text{geodesic}\right)
}\left( \tau \right) \right] _{\text{cor.}}$ must be positive by
construction, it must be the case that $A_{1}A_{2}<0$. Without loss of
generality, we assume $A_{1}=-A_{2}\equiv a\in 
\mathbb{R}
$. As a side remark, we point out that from (\ref{gsol1}) it follows that an
increase (decrease) in $\left\vert \mu _{x}\left( \tau \right) \right\vert $
is followed by an increase (decrease) in $\left\vert \mu _{y}\left( \tau
\right) \right\vert $. Therefore, internal consistency (consistency with the
class of geodesic paths considered) requires that we limit our analysis to
positively correlated microvariables, that is $r\in \left( 0\text{, }%
1\right) $. Finally\textbf{,} the average dynamical statistical volume $%
\widetilde{\emph{vol}}\left[ \mathcal{D}_{\Theta }^{\left( \text{geodesic}%
\right) }\left( \tau \right) \right] _{\text{cor.}}$ becomes,%
\begin{equation}
\widetilde{\emph{vol}}\left[ \mathcal{D}_{\Theta }^{\left( \text{geodesic}%
\right) }\left( \tau \right) \right] _{\text{cor.}}=\frac{1}{\tau }%
\int_{0}^{\tau }\emph{vol}\left[ \mathcal{D}_{\Theta }^{\left( \text{geodesic%
}\right) }\left( \tau ^{\prime }\right) \right] _{\text{cor.}}d\tau ^{\prime
}=\frac{a^{2}}{2\sigma _{0}}\frac{g^{\prime \frac{1}{2}}\left( r\right) }{%
\mathcal{A}^{\frac{3}{2}}\left( r\right) }\left[ \frac{1-\exp \left( -\sigma
_{0}\sqrt{\mathcal{A}\left( r\right) }\tau \right) }{\tau }\right] \text{.}
\end{equation}%
In the \emph{long-time limit}, the asymptotic behavior of the IGC becomes,%
\begin{equation}
\widetilde{\emph{vol}}\left[ \mathcal{D}_{\Theta }^{\left( \text{geodesic}%
\right) }\left( \tau \right) \right] _{\text{cor.}}\approx \frac{a^{2}}{%
2\sigma _{0}}\frac{g^{\prime \frac{1}{2}}\left( r\right) }{\mathcal{A}^{%
\frac{3}{2}}\left( r\right) }\frac{1}{\tau }\text{.}  \label{pd}
\end{equation}%
Thus, comparing the asymptotic expressions of the IGCs in the presence and
absence of microcorrelations, we obtain,%
\begin{equation}
\frac{\widetilde{\emph{vol}}\left[ \mathcal{D}_{\Theta }^{\left( \text{%
geodesic}\right) }\left( \tau \text{; }r\right) \right] _{\text{cor.}}}{%
\widetilde{\emph{vol}}\left[ \mathcal{D}_{\Theta }^{\left( \text{geodesic}%
\right) }\left( \tau \text{; }0\right) \right] _{\text{no-cor.}}}=\frac{%
g^{\prime \frac{1}{2}}\left( r\right) }{\mathcal{A}^{\frac{3}{2}}\left(
r\right) }\frac{\mathcal{A}^{\frac{3}{2}}\left( 0\right) }{g^{\prime \frac{1%
}{2}}\left( 0\right) }=\frac{1}{2^{\frac{5}{2}}}\sqrt{\frac{4\left(
4-r^{2}\right) }{\left( 2-2r^{2}\right) ^{2}}}\left( \frac{2+r}{4\left(
1-r^{2}\right) }\right) ^{-\frac{3}{2}}\overset{\text{def}}{=}\mathcal{F}_{%
\mathcal{M}_{S}}\left( r\right) \text{.}  \label{ratio}
\end{equation}%
Written alternatively\textbf{,}%
\begin{equation}
\widetilde{\emph{vol}}\left[ \mathcal{D}_{\Theta }^{\left( \text{geodesic}%
\right) }\left( \tau \text{; }r\right) \right] _{\text{cor.}}=\left[ \frac{1%
}{2^{\frac{5}{2}}}\sqrt{\frac{4\left( 4-r^{2}\right) }{\left(
2-2r^{2}\right) ^{2}}}\left( \frac{2+r}{4\left( 1-r^{2}\right) }\right) ^{-%
\frac{3}{2}}\right] \cdot \widetilde{\emph{vol}}\left[ \mathcal{D}_{\Theta
}^{\left( \text{geodesic}\right) }\left( \tau \text{; }0\right) \right] _{%
\text{no-cor.}}\text{.}  \label{KEY-RESULT}
\end{equation}%
We emphasize that $\mathcal{F}_{\mathcal{M}_{S}}\left( r\right) $\textbf{\ }%
is a monotonically decreasing function of\textbf{\ }$r$, that is $\mathcal{F}%
_{\mathcal{M}_{S}}\left( r_{1}\right) \geq \mathcal{F}_{\mathcal{M}%
_{S}}\left( r_{2}\right) $ for any $r_{1}\leq r_{2}$ with $r_{1}$, $r_{2}\in
\left( 0\text{, }1\right) $ and\textbf{\ }$0\leq \mathcal{F}_{\mathcal{M}%
_{S}}\left( r\right) \leq 1$ for $r\in \left( 0\text{, }1\right) $. We
observe an asymptotic power law decay of the IGC in (\ref{pd}) at a rate
determined by the correlation coefficient $r$. The ratio between the IGC in
the presence and in the absence of microcorrelations in (\ref{ratio}) leads
to conclude that microcorrelations cause an asymptotic information geometric
compression of the explored statistical macrostates at a\textbf{\ }faster
rate than the that observed in absence of microcorrelations. Our finding
presented in (\ref{KEY-RESULT}) shows an important connection between
(micro)-correlations and (macro)-complexity in Gaussian statistical models.

\section{Final Remarks}

In this article, we presented an analytical computation of the asymptotic
temporal behavior of the IGC for a finite-dimensional microcorrelated
Gaussian statistical model. The ratio between the IGC in the presence and in
absence of microcorrelations was explicitly computed. We observed a power
law decay of the IGC at a rate determined by the correlation coefficient.
Specifically, the presence of microcorrelations lead to the emergence of an
asymptotic information geometric compression of the statistical macrostates
explored by the system at a faster rate than that observed in absence of
microcorrelations. This result constitute an important and \emph{explicit}
connection between (micro)-correlations and (macro)-complexity in
statistical dynamical systems. The relevance of our finding is twofold:
first, it provides a neat description of the effect of information encoded
in \emph{microscopic} variables on experimentally observable quantities
defined in terms of dynamical \emph{macroscopic} variables \cite{carlo-MPLB}%
; second, it clearly shows the change in behavior of the macroscopic \emph{%
complexity} of a statistical model caused by the existence of \emph{%
correlations} at the underlying microscopic level.

We are confident that this work constitutes an important preliminary step
towards the computation of the asymptotic behavior of the dynamical
complexity of microscopically correlated multidimensional Gaussian
statistical models and other models of relevance in more realistic physical
systems.\textbf{\ }In principle, our approach extends its application to
arbitrary statistical models that may arise upon maximization of the
logarithmic relative entropy subject to the selected relevant information
constraints. In particular, our findings here presented could find practical
applications in the statistical analysis of biological and social systems
since Gaussian statistical models are of primary importance in statistical
studies \cite{har75}. However\textbf{,} our ultimate hope is to extend this
approach in the field of Quantum Information to better understand the
connection between quantum correlations (entanglement) and quantum
complexity \cite{nielsen, carlo-MPLB, cafaroPA, prosen, benenti}.

\begin{acknowledgments}
C. C. thanks A. Giffin and C. Lupo for useful discussions. The work of C. C.
and S. M. was supported by the European Community's Seventh Framework
Program (\emph{CORNER Project}; FP7/2007-2013) under grant agreement 213681.
D.-H. K. was supported by the National Research Foundation of Korea (NRF),
grant funded by the Korea government (MEST) through the Center for Quantum
Spacetime (CQUeST) of Sogang University, grant number 2005-0049409; D.-H. K.
also acknowledges the support of WCU (World Class University) program of
NRF/MEST (R32-2009-000-10130-0).
\end{acknowledgments}

\appendix

\section{Integration of the Geodesic Equations}

From the first and second\textbf{\ }Equations in (\ref{GEE}), we obtain,%
\begin{equation}
\frac{\ddot{\mu}_{x}\left( \tau \right) }{\dot{\mu}_{x}\left( \tau \right) }%
=2\frac{\dot{\sigma}\left( \tau \right) }{\sigma \left( \tau \right) }\text{
and }\frac{\ddot{\mu}_{y}\left( \tau \right) }{\dot{\mu}_{y}\left( \tau
\right) }=2\frac{\dot{\sigma}\left( \tau \right) }{\sigma \left( \tau
\right) }\text{,}  \label{ge1}
\end{equation}%
respectively. From (\ref{ge1}) it follows that,%
\begin{equation}
\dot{\mu}_{x}\left( \tau \right) =A_{1}\sigma ^{2}\left( \tau \right) \text{
and }\dot{\mu}_{y}\left( \tau \right) =A_{2}\sigma ^{2}\left( \tau \right) 
\text{,}  \label{s1}
\end{equation}%
where $A_{1}$ and $A_{2}$ are \emph{real} constants. Substituting (\ref{s1})
in the third Equation of (\ref{GEE}) we obtain,%
\begin{equation}
\ddot{\sigma}\left( \tau \right) \sigma \left( \tau \right) -\dot{\sigma}%
^{2}\left( \tau \right) +\frac{A_{1}^{2}+A_{2}^{2}-rA_{1}A_{2}}{4\left(
1-r^{2}\right) }\sigma ^{4}\left( \tau \right) =0\text{.}
\end{equation}%
Therefore, the coupled system of differential equations reduces to,%
\begin{eqnarray}
\dot{\mu}_{x}\left( \tau \right) -A_{1}\sigma ^{2}\left( \tau \right) &=&0%
\text{,}  \notag \\
&&  \notag \\
\dot{\mu}_{y}\left( \tau \right) -A_{2}\sigma ^{2}\left( \tau \right) &=&0%
\text{,}  \notag \\
&&  \notag \\
\ddot{\sigma}\left( \tau \right) \sigma \left( \tau \right) -\dot{\sigma}%
^{2}\left( \tau \right) +\mathcal{A}\left( r\right) \sigma ^{4}\left( \tau
\right) &=&0\text{,}
\end{eqnarray}%
where we recall that,%
\begin{equation}
\mathcal{A}\left( r\right) \overset{\text{def}}{=}\frac{%
A_{1}^{2}+A_{2}^{2}-rA_{1}A_{2}}{4\left( 1-r^{2}\right) }\text{.}
\label{adef}
\end{equation}%
We now proceed as follows: integrate the nonlinear differential equation $%
\ddot{\sigma}\left( \tau \right) \sigma \left( \tau \right) -\dot{\sigma}%
^{2}\left( \tau \right) +\mathcal{A}\left( r\right) \sigma ^{4}\left( \tau
\right) =0$ and then calculate $\mu _{x}\left( \tau \right) $ and $\mu
_{y}\left( \tau \right) $.

Letting $y\left( \tau \right) \overset{\text{def}}{=}\sigma \left( \tau
\right) $, the first nonlinear differential equation to integrate becomes, 
\begin{equation}
\ddot{y}\left( \tau \right) y\left( \tau \right) -\dot{y}^{2}\left( \tau
\right) +\mathcal{A}\left( r\right) y^{4}\left( \tau \right) =0\text{.}
\label{nde}
\end{equation}%
Performing the following change of variables,%
\begin{equation}
y\left( \tau \right) =\frac{dx\left( \tau \right) }{d\tau }=\dot{x}\left(
\tau \right)
\end{equation}%
equation (\ref{nde}) becomes%
\begin{equation}
\dot{x}\dddot{x}-\ddot{x}^{2}+\mathcal{A}\left( r\right) \dot{x}^{4}=0\text{.%
}  \label{nde2}
\end{equation}%
Equation (\ref{nde2}) can be integrated as follows. Performing the following
additional change of variables,%
\begin{equation}
\dot{x}=\frac{dx\left( \tau \right) }{d\tau }=z\left( x\right)  \label{a}
\end{equation}%
leads to%
\begin{equation}
\ddot{x}=zz^{\prime }\text{ and, }\dddot{x}=\left( z^{\prime \prime
}z+z^{\prime 2}\right) z\text{,}  \label{b}
\end{equation}%
with $z^{\prime }=\frac{dz}{dx}$. Substituting (\ref{a}) and (\ref{b}) into (%
\ref{nde2}), we find%
\begin{equation}
z^{\prime \prime }+\mathcal{A}\left( r\right) z=0\text{.}  \label{nde3}
\end{equation}%
Integration of (\ref{nde3}) yields%
\begin{equation}
z\left( x\right) =A_{3}\cos \left( \sqrt{\mathcal{A}\left( r\right) }%
x\right) +A_{4}\sin \left( \sqrt{\mathcal{A}\left( r\right) }x\right) \text{,%
}
\end{equation}%
where $A_{3}$ and $A_{4}$ are \emph{real} constants. Recalling that $\dot{x}=%
\frac{dx\left( \tau \right) }{dt}=z\left( x\right) $, we have%
\begin{equation}
\int \frac{dx}{A_{3}\cos \left( \sqrt{\mathcal{A}\left( r\right) }x\right)
+A_{4}\sin \left( \sqrt{\mathcal{A}\left( r\right) }x\right) }=\int d\tau
+A_{5}\text{.}  \label{sup}
\end{equation}%
\emph{Reality} conditions imply that $A_{3}=0$ and without loss of
generality we can set $A_{5}=0$. Integration of (\ref{sup})\ leads to,%
\begin{equation}
x\left( \tau \right) =\frac{1}{\sqrt{\mathcal{A}\left( r\right) }}\arccos
\left( \frac{1-\exp \left( 2A_{4}\sqrt{\mathcal{A}\left( r\right) }\tau
\right) }{1+\exp \left( 2A_{4}\sqrt{\mathcal{A}\left( r\right) }\tau \right) 
}\right) \text{.}
\end{equation}%
Finally, recalling that $y\left( \tau \right) =\frac{dx\left( \tau \right) }{%
d\tau }$ and $y\left( \tau \right) \overset{\text{def}}{=}\sigma \left( \tau
\right) $, we get%
\begin{equation}
\sigma \left( \tau \right) =2A_{4}\frac{\exp \left( A_{4}\sqrt{\mathcal{A}%
\left( r\right) }\tau \right) }{1+\exp \left( 2A_{4}\sqrt{\mathcal{A}\left(
r\right) }\tau \right) }\text{.}  \label{sigma}
\end{equation}%
Note that $\sigma \left( \tau \right) $ in (\ref{sigma}) satisfies the
equation $\ddot{\sigma}\left( \tau \right) \sigma \left( \tau \right) -\dot{%
\sigma}^{2}\left( \tau \right) +\mathcal{A}\left( r\right) \sigma ^{4}\left(
\tau \right) =0$. Once we have obtained $\sigma \left( \tau \right) $, we
have%
\begin{equation}
\mu _{x}\left( \tau \right) =\int A_{1}\sigma ^{2}\left( \tau \right) d\tau
+A_{6}\text{ and }\mu _{y}\left( \tau \right) =\int A_{2}\sigma ^{2}\left(
\tau \right) d\tau +A_{7}\text{,}
\end{equation}%
where $A_{6}$ and $A_{7}$ are \emph{real} constants. Integrating, we get%
\begin{equation}
\mu _{x}\left( \tau \right) =-\frac{2A_{4}A_{1}}{\sqrt{\mathcal{A}\left(
r\right) }}\frac{1}{1+\exp \left( 2A_{4}\sqrt{\mathcal{A}\left( r\right) }%
\tau \right) }+A_{6}\text{,}
\end{equation}%
and%
\begin{equation}
\mu _{y}\left( \tau \right) =-\frac{2A_{4}A_{2}}{\sqrt{\mathcal{A}\left(
r\right) }}\frac{1}{1+\exp \left( 2A_{4}\sqrt{\mathcal{A}\left( r\right) }%
\tau \right) }+A_{7}\text{.}
\end{equation}%
Assuming the following boundary conditions $\sigma \left( 0\right) =\sigma
_{0}>0$, $\mu _{x}\left( \tau _{\infty }\right) =\mu _{y}\left( \tau
_{\infty }\right) =0$, we find that $\sigma _{0}=A_{4}$, $A_{6}=A_{7}=0$.
Finally, the geodesic trajectories become,%
\begin{eqnarray}
\mu _{x}\left( \tau \text{; }r\right) &=&-\frac{2\sigma _{0}A_{1}}{\sqrt{%
\mathcal{A}\left( r\right) }}\frac{1}{1+\exp \left( 2\sigma _{0}\sqrt{%
\mathcal{A}\left( r\right) }\tau \right) }\text{, }\mu _{y}\left( \tau \text{%
; }r\right) =-\frac{2\sigma _{0}A_{2}}{\sqrt{\mathcal{A}\left( r\right) }}%
\frac{1}{1+\exp \left( 2\sigma _{0}\sqrt{\mathcal{A}\left( r\right) }\tau
\right) }\text{, }  \notag \\
&&  \notag \\
\sigma \left( \tau \text{; }r\right) &=&2\sigma _{0}\frac{\exp \left( \sigma
_{0}\sqrt{\mathcal{A}\left( r\right) }\tau \right) }{1+\exp \left( 2\sigma
_{0}\sqrt{\mathcal{A}\left( r\right) }\tau \right) }\text{,}  \label{gsol}
\end{eqnarray}%
with $\mathcal{A}\left( r\right) $ defined in (\ref{adef}), $A_{1}$ and $%
A_{2}$ \emph{real} constants and $\sigma _{0}>0$. Note that $\sigma \left(
\tau \text{; }r\right) \in \left( 0\text{, }+\infty \right) $ while $\mu
_{x}\left( \tau \text{; }r\right) $ and $\mu _{y}\left( \tau \text{; }%
r\right) \in \left( -\infty \text{, }+\infty \right) $.


\begin{thebibliography}{99}
\bibitem{gell-mann} M. Gell-Mann, "\emph{What is Complexity}", Complexity 
\textbf{1}, 1 (1995).

\bibitem{caticha1} A. Caticha, "\emph{Entropic Dynamics}", in \textit{%
Bayesian Inference and Maximum Entropy Methods in Science and Engineering},
ed. by R.L. Fry, AIP Conf. Proc. \textbf{617}, 302 (2002).

\bibitem{carlo-tesi} C. Cafaro, "\emph{The Information Geometry of Chaos}",
Ph. D. Thesis, SUNY at Albany, NY-USA (2008).

\bibitem{carlo-CSF} C. Cafaro, "\emph{Works on an information
geometrodynamical approach to chaos}", Chaos, Solitons \& Fractals \textbf{41%
}, 886 (2009).

\bibitem{caticha2} A. Caticha and R. Preuss, "\emph{Maximum entropy and
Bayesian data analysis: Entropic prior distributions}", Phys. Rev. \textbf{%
E70}, 046127 (2004).

\bibitem{adom1} A. Giffin, "\emph{Maximum Entropy: The Universal Method for
Inference}", Ph. D. Thesis, SUNY at Albany, NY-USA (2008).

\bibitem{amari} S. Amari and H. Nagaoka, \emph{Methods of Information
Geometry}, American Mathematical Society, Oxford University Press, 2000.

\bibitem{casetti} L. Casetti, C. Clementi, and M. Pettini, "\emph{Riemannian
theory of Hamiltonian chaos and Lyapunov exponents}", Phys. Rev. \textbf{E54}%
, 5969 (1996).

\bibitem{di bari} M. Di Bari and P. Cipriani, "$\emph{Geometry}$ $\emph{and}$
$\emph{Chaos}$ $\emph{on}$ $\emph{Riemann}$ $\emph{and}$ $\emph{Finsler}$ $%
\emph{Manifolds}$", Planet. Space Sci. \textbf{46}, 1543 (1998).

\bibitem{jacobi} C. G. J. Jacobi, "\emph{Vorlesungen uber Dynamik}", Reimer,
Berlin (1866).

\bibitem{kawabe} T. Kawabe, "\emph{Indicator of chaos based on the
Riemannian geometric approach}", Phys. Rev. \textbf{E71}, 017201 (2005); T.
Kawabe, "\emph{Chaos based on Riemannian geometric approach to Abelian-Higgs
dynamical system}", Phys. Rev. \textbf{E67}, 016201 (2003).

\bibitem{leb81} J. L. Lebowitz, "\emph{Microscopic Dynamics and Macroscopic
Laws}", Annals of the New York Academy of Sciences \textbf{373}, 220 (1981).

\bibitem{leb93} J. L. Lebowitz, "\emph{Macroscopic Laws, Microscopic
Dynamics, Time's Arrow and Boltzmann's Entropy}", Physica \textbf{A194}, 1
(1993).

\bibitem{leb99} J. L. Lebowitz, "\emph{Microscopic Origins of Irreversible
Macroscopic Behavior}", Physica \textbf{A263}, 516 (1999).

\bibitem{tom90} T. Toffoli, "\emph{How Cheap Can Mechanics' First Principles
Be?}", in "Complexity, Entropy and The Physics of Information" edited by W.
H. Zurek, page 301, Addison-Wesley (1990).

\bibitem{shib99} T. Shibata et \textit{al}., "\emph{Noiseless Collective
Motion out of Noisy Chaos}", Phys. Rev. Lett. \textbf{82}, 4424 (1999).

\bibitem{tribus} M. Tribus\textbf{, "}\emph{Rational Descriptions,\
Decisions and Designs}\textbf{", }Pergamon Press (1969)\textbf{.}

\bibitem{caticha(REII)} A. Caticha, "\emph{Relative Entropy and Inductive
Inference}", \textit{Bayesian Inference and Maximum Entropy Methods in
Science and Engineering},ed. by G. Erickson and Y. Zhai, AIP Conf. Proc.%
\textbf{\ 707}, 75 (2004).

\bibitem{caticha-giffin} A. Caticha and A. Giffin, "\emph{Updating
Probabilities}", in \textit{Bayesian Inference and Maximum Entropy Methods
in Science and Engineering, ed.} by Ali Mohammad-Djafari, AIP Conf. Proc. 
\textbf{872}, 31 (2006).

\bibitem{jay57} E. T. Jaynes, "\emph{Information theory and statistical
mechanics, I}", Phys. Rev. \textbf{106}, 620 (1957); \ "\emph{Information
theory and statistical mechanics, II}", Phys. Rev. \textbf{108}, 171 (1957).

\bibitem{cafaroIJTP} C. Cafaro, \textquotedblleft \emph{%
Information-Geometric Indicators of Chaos in Gaussian Models on Statistical
Manifolds of Negative Ricci Curvature}\textquotedblright , Int. J. Theor.
Phys. \textbf{47}, 2924 (2008).

\bibitem{roz} Y. A. Rozanov, "\emph{Probability Theory: A Concise Course}",
Dover Publications, New York (1977).

\bibitem{felice} F. De Felice and J. S. Clarke, "\emph{Relativity on Curved
Manifolds}", Cambridge University Press (1990).

\bibitem{cafaroPD} C. Cafaro and S. A. Ali, "\emph{Jacobi Fields on
Statistical Manifolds of Negative Curvature}", Physica \textbf{D234}, 70
(2007).

\bibitem{har75} D. H. Hartnett, "\emph{Introduction to Statistical Methods}%
", Addison-Wesley (1975).

\bibitem{nielsen} M. A. Nielsen, "\emph{Quantum information science as an
approach to complex quantum systems}", arXiv:quant-ph/0208078 (2002).

\bibitem{carlo-MPLB} C. Cafaro, "\emph{Information geometry, inference
methods and chaotic energy levels statistics}", Mod. Phys. Lett. \textbf{B22}%
, 1879 (2008).

\bibitem{cafaroPA} C. Cafaro and S. A. Ali, "\emph{Can chaotic quantum
energy levels statistics be characterized using information geometry and
inference methods?}", Physica \textbf{A387}, 6876 (2008).

\bibitem{prosen} T. Prosen, "\emph{Chaos and Complexity of Quantum Motion}",
J. Phys. \textbf{A40}, 7881 (2007).

\bibitem{benenti} G. Benenti and G. Casati, "\emph{How complex is quantum
motion?}", Phys. Rev \textbf{E79}, 025201 (2009).
\end{thebibliography}
\end{document}